\documentclass[pre,onecolumn,superscriptaddress]{revtex4}

\usepackage{amssymb,amsbsy}
\usepackage{graphicx}
\usepackage{rotating}
\usepackage{dcolumn}
\usepackage{amsmath}

\newcommand{\av}[1]{\vert #1 \vert}

\begin{document}
\title{Mixing Patterns in Interdisciplinary Collaboration Networks: Assessing Interdisciplinarity Through Multiple Lenses}

\author{Shihui Feng}
\affiliation{Unit of Human Communication, Development, and Information Sciences, The University of Hong Kong, Hong Kong}

\author{Alec Kirkley}
\affiliation{Department of Physics, University of Michigan, Ann Arbor, USA}

\begin{abstract}

There are inherent challenges to interdisciplinary research collaboration, such as bridging cognitive gaps and balancing transaction costs with collaborative benefits. This raises the question: Does interdisciplinary research necessarily result in interdisciplinary collaborations? This study aims to explore this question and assess collaboration preferences in interdisciplinary research at the individual, dyadic, and team level by examining mixing patterns in a collaboration network. Using a network of over 2,000 researchers from the field of artificial intelligence in education, we find that ``interdisciplinarity” is demonstrated by diverse research experiences of individual researchers rather than diversity among researchers within collaborations. We also examine intergroup mixing by applying a novel approach to classify the active and non-active researchers in the collaboration network based on participation in multiple teams. We find a significant difference in indicators of academic performance and experience between the clusters of active and non-active researchers, suggesting intergroup mixing as a key factor in academic success. Our results shed light on the nature of team formation in interdisciplinary research, as well as highlight the importance of interdisciplinary programs.

\end{abstract}
\maketitle

\section{Introduction}

There are many significant social and global problems that cross disciplinary
boundaries. The scientific complexity of these problems calls for the synthesis of
concepts, theories and methods from multiple disciplines, and new research areas
beyond traditional disciplinary frameworks. Thus, the boundary of research disciplines
has been expanding, merging, and transforming dynamically, and there has been an
increasing trend towards more interdisciplinary research in both the natural and social
sciences since the mid-1980s \cite{van2015interdisciplinary}. In particular, with exponentially growing
digital data in various fields, formulating data-informed decisions requires both
subject domain expertise, as well as fluency with computational techniques to
process, analyze and interpret this large-scale data. Given the demand for diverse
skillsets required in interdisciplinary research, interdisciplinary collaboration is seen
as a means to extend the cognitive capability of a research team while handling
complex problems.\\

However, there are some obstacles in developing interdisciplinary collaboration
that cannot be neglected. Firstly, collaboration requires a common ground where a
group of individual researchers have a certain level of shared understanding and
mutual knowledge of the research problems \cite{clark1991grounding,hertzum2008collaborative}. In particular, social and natural scientists may have different perspectives and
approaches to defining, solving and presenting problems, which introduces philosophical
obstacles in interdisciplinary collaboration \cite{campbell2005overcoming}. Committing to an interdisciplinary collaboration poses a risk for researches from different disciplines, in terms of the balance of transaction costs and collaborative
benefits. The motivation for researchers to participate in interdisciplinary
collaboration could highly depend on the evaluation of the perceived risks and
rewards. Secondly, obstacles relevant to psychosocial and practical perspectives
hindering collaboration in general can also be applied to interdisciplinary
collaboration. For instance, the “obstructive misconceptions or prejudices” between
social and natural scientists \cite{campbell2005overcoming} could result in a
lack of appreciation of each other’s value and contributions in collaboration, which
could influence the effectiveness and continuity of interdisciplinary collaboration.\\

Given the challenges of interdisciplinary collaboration, an essential question is
whether or not there is data-based evidence of homophily or diversity in
interdisciplinary research collaboration. This study aims to assess the
interdisciplinarity in research collaboration at the individual, dyadic, and team level
for research on artificial intelligence in education (AIED). AIED has been developing
fast as an interdisciplinary research area in the last decade, focusing on applying
computational techniques in analyzing large-scale educational data and developing
intelligent systems for supporting teaching and learning activities. It is a
demonstration of the newly emerging interdisciplinary research paradigm of
integrating computational science into social and humanities contexts. The mixing
patterns of a collaboration network in AIED research is studied following four
research questions: 1) Do individual researchers tend to have experience in multiple
disciplines? 2) Do researchers in an interdisciplinary area prefer to collaborate with
others from a similar or different research background? 3) Do teams as a whole tend
to be composed of researchers with similar, or diverse research backgrounds? 4) Do
researchers with different structural characteristics in a collaboration network have
different research performance? Our findings provide data-informed evidence for
understanding the underlying mechanisms of the formation of collaboration in
interdisciplinary research. These results can further yield insights for formulating
strategies and training programs to facilitate effective collaboration in
interdisciplinary research.

\section{Related work}
\subsection*{\textbf{The Paradox of “Interdisciplinarity” in Interdisciplinary Collaboration}}

A variety of group and organizational theories provide theoretical underpinnings for the formation, dynamics and complexity of academic collaboration. The formation of team members in research is vital to the success and effectiveness of collaboration. Lewin’s group dynamics theory \cite{lewin1948resolving} suggests that the shared incentives among group members and task interdependence significantly affect the group process in a collaboration, and places higher priority on the shared incentives rather than the similarity or dissimilarity of individuals. However, a certain level of similarity in characteristics of individuals could positively affect the development of shared commitment towards a goal. Ruef, Aldrich, and Carter \cite{ruef2003structure} provided supporting evidence that homophily, together with network ties, has the determining effects on group formation. Homophily in group composition refers to the tendency for people to collaborate with others who share a certain level of similarity on various attributes, for instance gender, age or ethnicity. In interdisciplinary collaboration, bringing individuals from different disciplines together introduces a heterogenous attribute to a group, which conflicts with the principle of homophily in group composition. Diverse academic backgrounds within a research team extend research capacity but also may increase the complexity and disequilibrium of group dynamics in collaboration. An essential question that comes along with this line of thinking is whether or not homophily is still an applicable mechanism for group composition in interdisciplinary research collaboration. \\

Previous studies are largely focused on studying the effects of interdisciplinary collaboration on professional practices in the context of healthcare \cite{baggs1992association,fewster2008interdisciplinary,petri2010concept}. Regarding the factors associated with the success of interdisciplinary collaboration, a study conducted by Van Rijnsoever and Hessels \cite{van2011factors} found that years of working experience, previous experience of working at other universities or firms, and being female are positively associated with interdisciplinary research collaboration. Cummings and Kiesler \cite{cummings2008collaborates} also found that prior collaboration experience plays an important role in eliminating the barriers in interdisciplinary collaboration. However, there is still a lack of research studying the group composition in interdisciplinary collaboration with a focus on the homogeneity or heterogeneity of team members’ research backgrounds. This study aims to address the novel question of assessing interdisciplinarity in interdisciplinary collaboration from individual, dyadic, and team levels using network approaches. 

\subsection*{\textbf{Network Approaches to Studying Research Collaboration}}
A network is a mathematical object from graph theory consisting of nodes connected in pairs by edges. Networks are a useful tool for representing pairwise relationships in various social or physical systems in an abstract manner. Consequently, network approaches have been widely applied to study the structure of relationships and interconnection among components within and across systems. Research collaboration can be well represented by networks consisting of researchers and the collaborative ties among them, and a large body of literature has studied research collaboration from a network science perspective. Guimera \cite{guimera2005team} studied the temporal structures of research collaboration networks and found that prior collaboration experience and the recruitment of newcomers has a positive effect on the success of research collaboration in multiple fields. Moody \cite{moody2004structure} analyzed the cohesion of research collaboration in sociology by examining a sociology collaboration network from 1963 to 1999. Dahlander and McFarland \cite{dahlander2013ties} identified six attributes of collaborative ties that affect the formation and persistence of research collaboration across time. In general, previous studies have primarily focused on the following aspects of collaboration networks: 1) Descriptive structural characteristics; 2) Group formation; 3) Temporal group dynamics; and 4) Structural factors associated with the success of collaboration. In this study, we focus on providing new insights about interdisciplinarity in collaboration networks through aspects (2) and (4) using novel measures and approaches. 

\section{Methodology}
\subsection{Data Collection}
\label{data}
The collaboration network data used in this study are collected from three representative journals on artificial intelligence in education (AIED), an emerging interdisciplinary research area. The three journals studied are \emph{International Journal of Artificial Intelligence in Education}, \emph{Proceedings of Educational Data Mining}, and \emph{Proceedings of Learning at Scales}. The bibliometric information of all the available publications from these journals during the years 2010 to 2019 are obtained from the DBLP database. The collaboration network is constructed with the 2022 authors in the dataset as nodes, with an edge between two nodes if these authors coauthored a paper together.\\

Previous studies define research disciplines of authors based on department affiliations \cite{qin1997types}. However, in interdisciplinary research areas, department affiliations are poor representations of an individual's research experience, as by the nature of the area, authors may not be easily classified by a single field. In this study, we propose a novel approach to defining the research disciplines of authors based on the discipline categories of their publications rather than their departmental affiliations. The Scopus database provides the number of papers per research field for indexed authors based on the groupings of journal discipline categories, which are extracted for each author in the dataset to represent the interdisciplinarity of their research background. Each author's publication counts were normalized to give the fraction of all of their work classified under a given category, which was represented with a vector with 27 entries, the number of unique disciplines for authors in the dataset. For example, if author $i$ has 50 publications classified under 'Computer Science', 30 publications classified under 'Math', and 20 publications classified under 'Sociology', they would have a vector $\vec{x}_i$ with entries $\{.5,.3,.2\}$ for the entries corresponding to these disciplines respectively, and $0$'s elsewhere.  \\  

Additionally, other author metadata is retrieved through the Scopus API, including their earliest and latest publication year, and h-index. We consider the research field with the highest number of publications of an author as their primary research discipline, but all the publication fields of an author are considered for assessing the interdisciplinarity of individual researchers. To explore the associations between structural properties of authors in the collaboration network and academic performance and experiences, the h-index is used as an indicator of academic success. Academic experience is measured based on the number of years between an author's first and latest publication.  

\subsection{Measures for Assessing Interdisciplinarity}
\label{measures}

Different measures were used to capture the interdisciplinarity of research collaboration at the individual, dyadic and team level, which we discuss here. In addition, we detail a simple scheme to classify active and non-active collaborators in the network based on their tie patterns, which allows us to explore the associations between research collaboration and academic performance and experiences. \\

\noindent\noindent \emph{Individual interdisciplinarity}: This refers to the diversity of the research fields that span an individual researcher's publication history, allowing us to address our first question of whether or not individual researchers tend to have experience in multiple disciplines. As it is an intuitive measure for the diversity of categorical data with clear upper and lower bounds \cite{kumar1985diversity}, entropy is used here to measure the variation of the fields comprising each individual researcher's publication history. Using the information from the publication count vector $\vec{x}_i$ collected in Sec. \ref{data}, the entropy for researcher $i$'s publication history is given by
\begin{align}
\label{ent}
H_i = -\frac{1}{\log(N_d)}\sum_{d=1}^{N_d}\vec{x}_{id}\log(\vec{x}_{id}),  \end{align}
where $\vec{x}_{id}$ is the fraction of researcher $i$'s publications classified under field $d$ (the $d$-th entry in the normalized publication count vector $\vec{x}_i$), and $N_d$ is the number of unique disciplines in the dataset (here, $N_d=27$). The prefactor $\log(N_d)^{-1}$ is to ensure that the entropy value lies between $0$ and $1$, which allows us to assess how high the entropy of a researcher's publication distribution is relative to its maximum possible value. Authors with only one publication field (1.2\% of all authors) are excluded in the analysis, as in this case $H_i$ is trivially $0$. High values of this index ($H_i$  close to $1$) indicate researchers with a high level of individual interdisciplinarity in their publication record, and low values ($H_i$ close to $0$) indicate researchers with a low level of interdisciplinarity. \\ 

\noindent\noindent \emph{Dyadic interdisciplinarity}: Dyadic interdisciplinarity assesses the level of similarity of research background for a pair of researchers in the collaboration network, addressing the second research question of whether or not homophily is still an applicable mechanism for explaining the collaboration preferences in interdisciplinarity research. For a surface level assessment of pairwise interdisciplinarity, the fraction of all edges that are comprised of researchers with the same primary discipline (the discipline in which an author published the most) is computed. However, to account for imbalances in the global distribution of primary affiliations (i.e. how many ties we expect between authors of the same primary discipline by chance), we compare this fraction with the same fraction computed on all pairs of authors who \emph{did not} collaborate. To see whether these fractions differ significantly, we use a two proportion $z$-test, the details of which we describe shortly. However, given the nature of interdisciplinary research, it is essential to take the diversity within each individual's research experience into consideration while assessing collaboration patterns, as individuals are not well categorized into a single research domain. We thus employ cosine similarity to measure the dyadic interdisciplinarity in the network, by comparing the publication count vectors for each of the authors. Cosine similarity is a common measure for determining the similarity of two non-zero vectors depending on their orientations in some high dimensional space, and in our context is given by
\begin{align}
\label{cos}
S_{ij} = \frac{\vec{x}_i\cdot \vec{x}_j}{\vert\vert \vec{x}_i \vert\vert \vert\vert \vec{x}_j \vert\vert},    
\end{align}
where $\vert\vert \vec{x}_i \vert\vert$ is the magnitude of $\vec{x}_i$. The value of $S_{ij}$ is also restricted to $[0,1]$, and a high value of $S_{ij}$ indicates a high similarity in the research backgrounds of authors $i$ and $j$, while a low value indicates dissimilarity. We also compute Eq. \ref{cos} for both edges and non-edges to see whether researchers collaborate with others that are more or less similar than themselves. \\

\noindent\noindent \emph{Team interdisciplinarity}: To address the third research question of whether research teams in an interdisciplinary area  tend to be composed as a whole of researchers with similar or diverse backgrounds, we look at team interdisciplinarity. This is also assessed based on both primary discipline and publication vectors $\vec{x}$ to give results from multiple perspectives. Within-group entropy is employed to assess the team interdisciplinarity based on the primary publication fields for all authors in a research group. In a similar manner to Eq. \ref{ent}, the within-group entropy $\tilde H_p$ for a paper $p$ is given by
\begin{align}
\label{ent_group_primary}
\tilde H_p = -\frac{1}{\log(\text{Max}\{\av{p},N_d\})}\sum_{d=1}^{N_d}f_{pd}\log(f_{pd}),
\end{align}
where $f_{pd}$ is the fraction of authors on the paper $p$ with primary discipline $d$, and $\av{p}$ is the number of authors on paper $p$. The new normalization factor $\log(\text{Max}\{\av{p},N_d\})^{-1}$ is introduced here because the upper bound on the entropy of collaboration $p$ is restricted by either the size of the collaboration or the number of possible disciplines (whichever is larger). Additionally, a similar manner to the analysis on dyadic interdisciplinarity, the within-group average cosine similarity is used to assess the team interdisciplinarity beyond looking simply at primary discipline. The mean within-group cosine similarity $\tilde S$ for paper $p$ is given by 
\begin{align}
\label{group_cos}
\tilde S_p = \frac{2}{\av{p}(\av{p}-1)}\sum_{(i,j)\in p}S_{ij},    
\end{align}
where the prefactor normalizes the measure to $[0,1]$, and the sum is over all pairs of nodes in $p$. The measures in Eq.s \ref{ent_group_primary} and \ref{group_cos} can be interpreted in a similar manner as the measures in Eq.s \ref{ent} and \ref{cos} respectively, except they assess team-level interdisciplinarity rather than individual or pairwise interdisciplinarity.  
\\

\noindent\noindent \emph{Core-shell decomposition}: The last research question examines the associations of the structural characteristics and academic performance and experience in the collaboration network. We define active collaborators in the network as researchers who are active in collaborating with multiple research groups in multiple projects. These authors published more than one article with diverse groups and perform a significant role in contributing to the global connectivity of research collaboration in the field, but may have a low level of local transitivity. Local transitivity, which we denote $C_i$ for an author $i$, refers to the fraction of all possible ties that exist among $i$'s neighbors, and is given by
\begin{align}
C_i = \frac{2}{\av{\partial_i}(\av{\partial_i}-1)}\sum_{(j,k)\in \partial_i}A_{jk}   
\end{align}
where $\partial_i$ is the neighborhood of $i$, and $A_{jk}$ is the binary adjacency matrix such that $A_{jk}=1$ if there is a connection between $i$ and $j$, and $A_{jk}=0$ if there is not. Collaboration networks constructed using co-authorship data tend to have a large number of fully connected cliques: co-authors of the same research paper are fully connected. Therefore, a high number of nodes have a maximum local transitivity ($C_i=1$), as they only collaborate with members of their research group. Thus, simply by looking for nodes $i$ with local clustering coefficient $C_i<1$, we can identify the nodes that act as bridges in the collaboration network by associating those with $C_i<1$ as the ``core" of the network and those with $C_i=1$ as the ``shell". In this way, we can see how the network separates into nodes with topologically diverse neighborhoods and nodes with homogeneous connectivity. There are other measures to assess the level of global connectivity a node facilitates (such as betweenness centrality), but here we are only concerned with a binary classification of whether a node is \emph{active} in collaboration (has multiple distinct groups of collaborators) or \emph{inactive} (has only one group of collaborators). As computation of local clustering coefficients is fast on most networks, this method is a relatively cost-effective approach for performing a decomposition of a collaboration network into a core and a shell. Removing nodes $i$ with $C_i=1$ and iteratively identifying nodes of $C_i=1$, we can decompose the network into nodes with different 'coreness' values, which gives a more sophisticated means of identifying the importance of nodes for the global connectivity of the network, but we leave this and other extensions to future work.\\

\section{Results}
\subsection{Individual Interdisciplinarity}
\begin{figure}
    \centering
    \includegraphics[width=.6\textwidth]{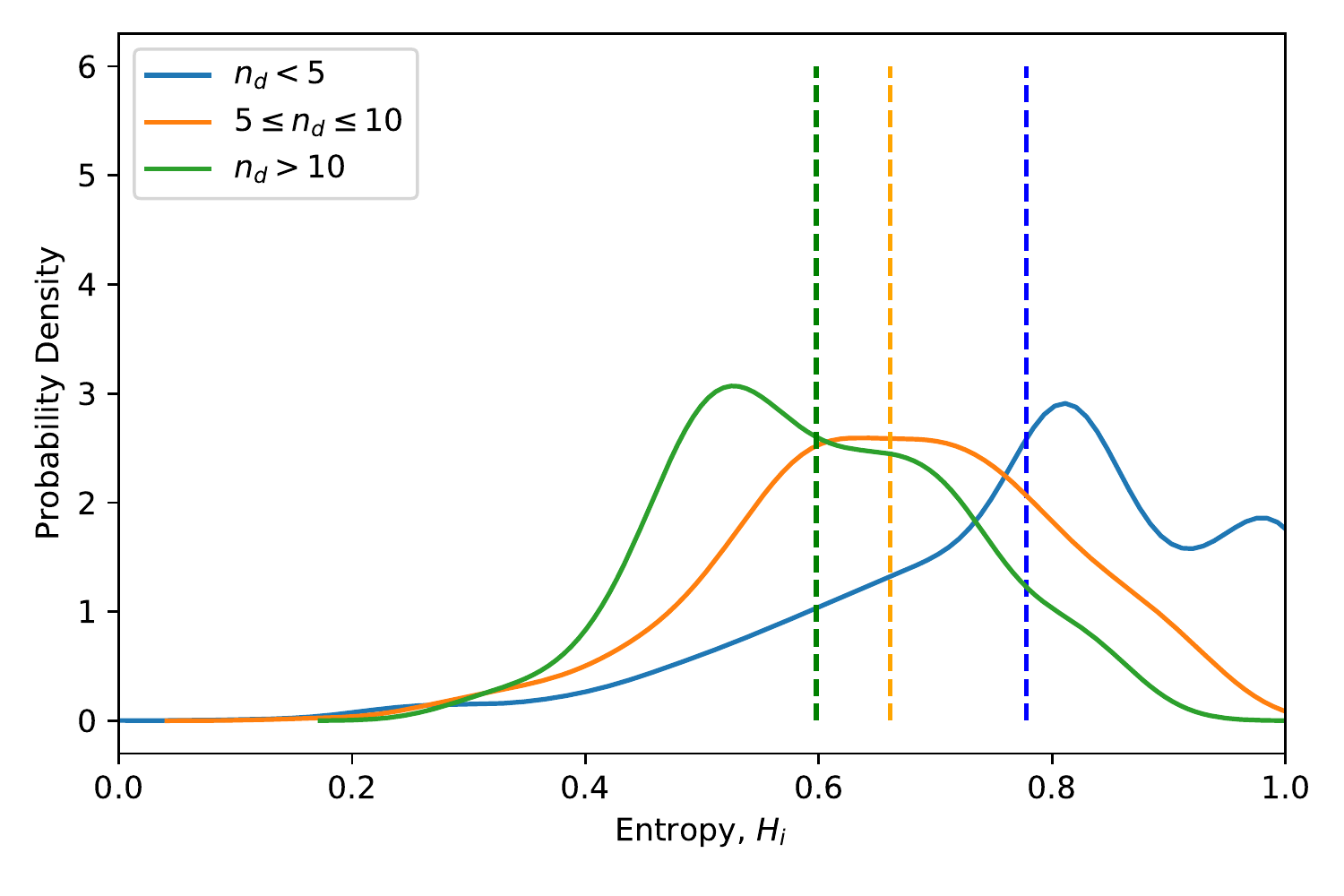}
    \caption{Probability densities of entropy (Eq. \ref{ent}) for researchers with varying numbers of publication disciplines, $n_d$, with the mean of each distribution indicated by a vertical line. Individual researchers had relatively high diversity in their research backgrounds, with those active in more fields showing a slightly less spread out research profile.}
    \label{ind_ent_fig}
\end{figure}

The 2022 authors in the collaboration network are from 18 primary disciplines and have research experiences in 27 disciplines in total. In Figure \ref{ind_ent_fig}, we plot the distribution of the entropies (Eq. \ref{ent}) for all researchers that contributed to a given number of subfields (i.e. had that many non-zero entries in $\vec{x}_i$). For easier visualization purposes, the histograms were smoothed using a kernel density estimate to obtain a probability density function. Based on the densities in the figure, we can see that authors in the interdisciplinarity area contribute relatively equally to all the fields they publish in ($H_i$ is moderately high on average), but that the distributions vary depending on how many fields an author participated in. In particular, authors with more publication fields are not able to contribute equally to all of these fields, and so we see a systematic decrease in the position of the $H_i$ values. The individual interdisciplinarity distributions for each individual journal all present similar results, and so the trends we see persist at the journal-level as well. \\

\subsection{Dyadic Interdisciplinarity}
\label{dyadic}
Among 5002 edges between the 2022 authors in the network, $81\%$ of pairs have the same primary discipline, while $75\%$ of non-edge pairs have the same primary discipline (a majority of authors have computer science as a primary discipline). Using a two-proportion $z$-test, this difference is statistically significant at the $1\%$ level ($z=11.7,\;p<.01$). These results suggest that authors preferentially collaborate with others of the same primary discipline. However, as discussed in Sec. \ref{measures}, we need to go beyond primary disciplines to analyze interdisciplinarity in an interdisciplinary research field, so cosine similarity (Eq. \ref{cos}) is also examined across all edge and non-edge pairs. Figure \ref{cos_fig} shows the probability densities of $S_{ij}$ over these pairs, indicating a shift in the distribution for edges towards higher similarity values than for the non-edges. We test the null hypothesis that it is equally likely that a randomly selected value from the edge distribution is less than or greater than a randomly selected value from the non-edge distribution using a Mann-Whitney U test, finding that we can reject this null in favor of the alternative hypothesis that the cosine similarities on the edges are systematically higher than on the non-edges (\text{Median 1}$=0.95$, \text{Median 2}$=0.89$, $n_1=5002$, $n_2=2.04\times 10^6$, $U \gg 10$ , $p \ll .01$, one-tailed). We also report the results from a Kolmogorov-Smirnov test to determine whether the distributions are the same, which also indicates significant differences between edges and non-edges ($D=0.22$, $p\ll .01$). Our findings suggest that, from a pairwise lens, researchers prefer to collaborate with those who have the same major discipline as themselves. Additionally, the results suggest that researchers prefer collaborators with similar interdisciplinary research backgrounds. The same analysis is also performed for all individual journal networks, and the results are consistent with the findings for the full network. We also plot the collaboration network, with edges colored from according the their cosine similarity $S_{ij}$, for visual inspection, in Figure \ref{edge_colors}. \\

\begin{figure}
    \centering
    \includegraphics[width=.6\textwidth]{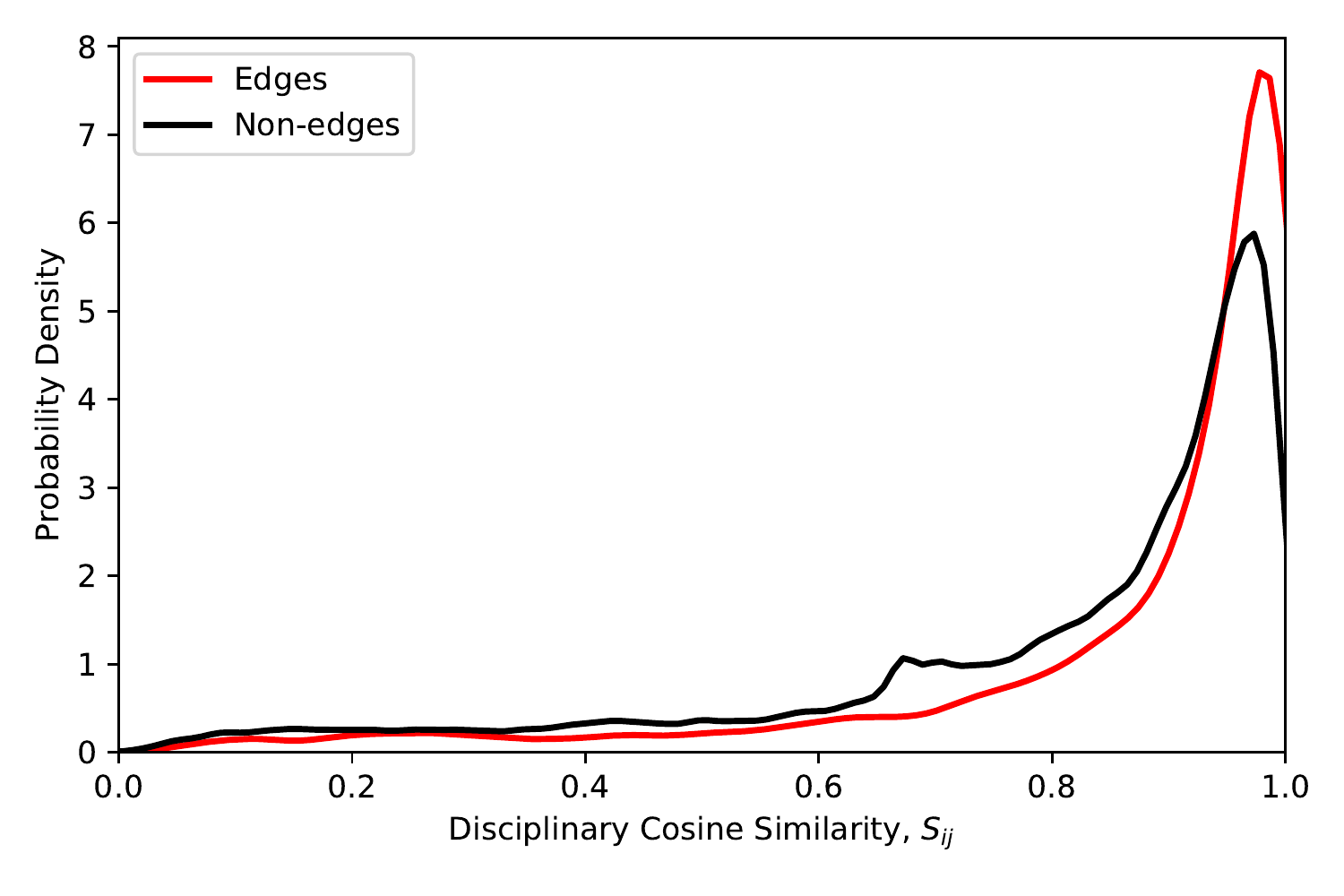}
    \caption{Probability densities of cosine similarity (Eq. \ref{cos}) for edges and non-edges. The $S_{ij}$ values for edges tended to be higher than those for non-edges, an effect that is shown to be statistically significant through Mann-Whitney and Kolmogorov-Smirnov tests.}
    \label{cos_fig}
\end{figure}

\begin{figure}
    \centering
    \includegraphics[width=.6\textwidth]{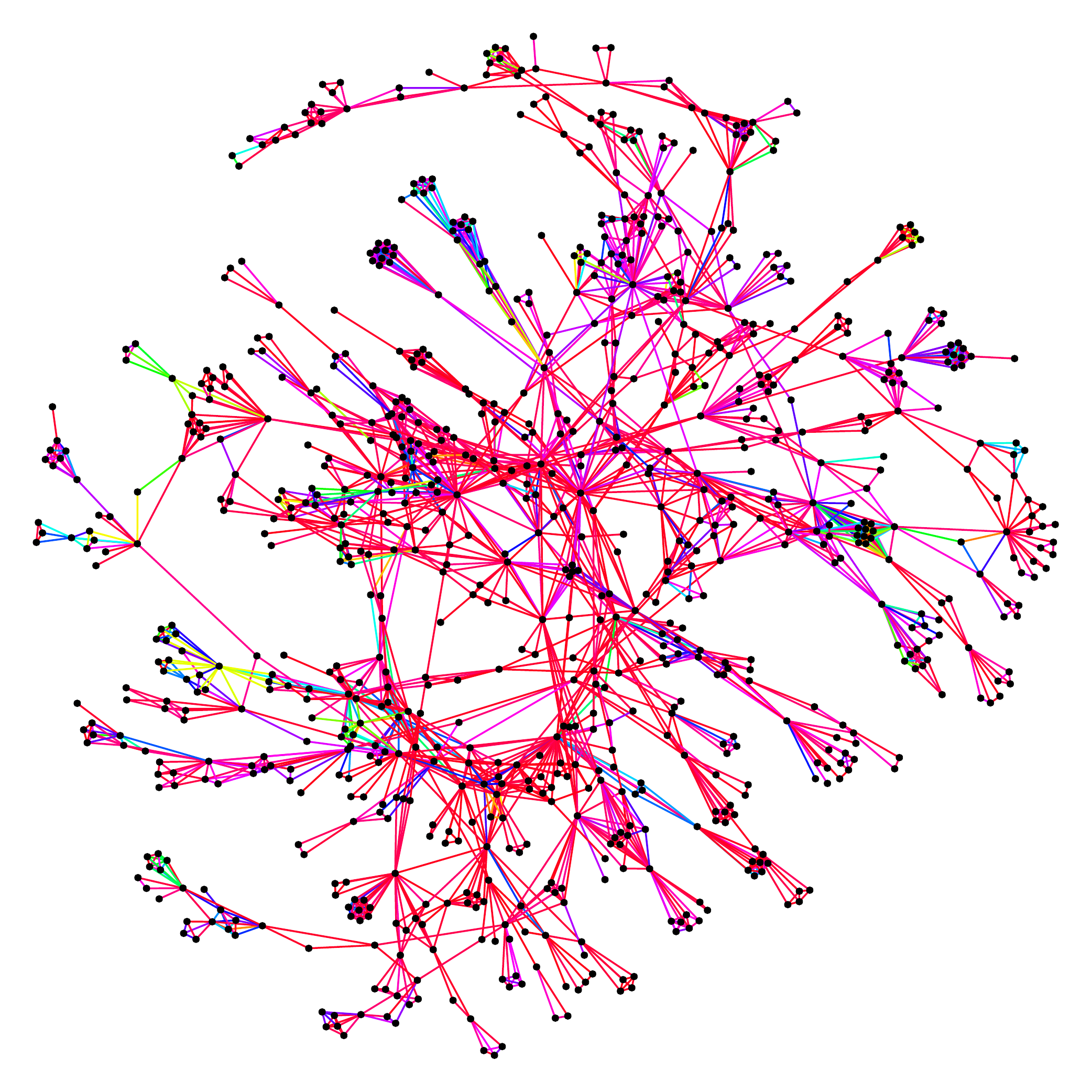}
    \caption{The giant component of the collaboration network used in this study, with edges colored according to interdisciplinary cosine similarity $S_{ij}$ (Eq. \ref{cos}). The values on the edges range from $S_{ij}=0$ (violet) to $S_{ij}=1$ (red) within the visible spectrum. Most edges have a high level of interdisciplinary similarity, as demonstrated in the analysis of Section \ref{dyadic}.}
    \label{cos_fig}
\end{figure}

\subsection{Team Interdisciplinarity}
\begin{figure}
    \centering
    \includegraphics[width=.6\textwidth]{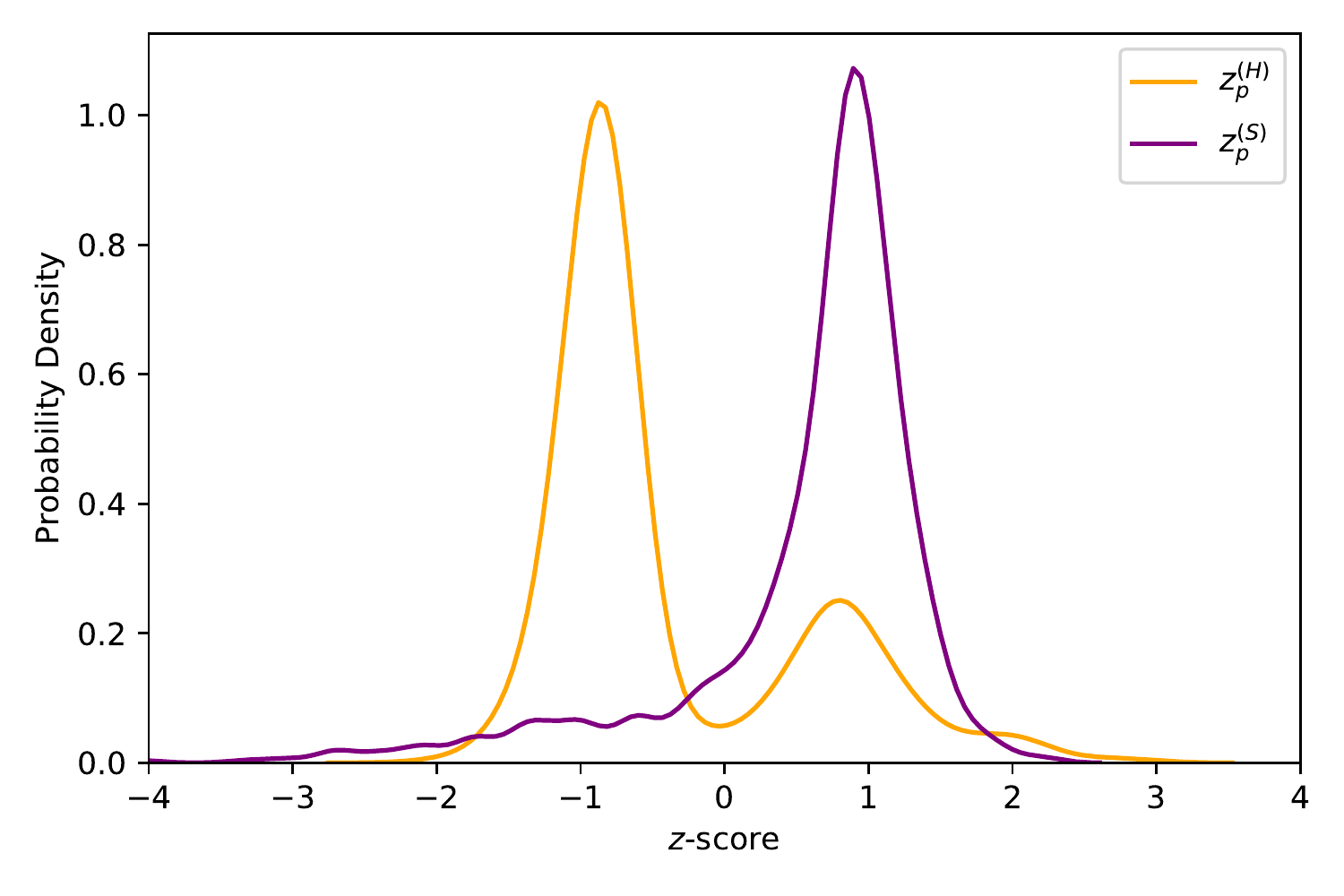}
    \caption{Probability densities of team interdisciplinarity $z$-scores in Equations \ref{zH} and \ref{zS} for the collaboration network. Both distributions suggest that research teams are more homogeneous than expected by chance.}
    \label{group_fig}
\end{figure}

To assess whether or not homophily is still an applicable mechanism to explain team formation in research collaboration, we examine interdisciplinarity at the team level. A group of co-authors of the same articles is considered a research team, represented as a fully connected clique in the collaboration network. To assess the interdisciplinarity of a team solely considering primary disciplines, we compute $\tilde H_p$ from Eq. \ref{ent_group_primary} on all the research teams $p$ in the network. We also compute Eq. \ref{ent_group_primary} on 1000 randomized teams (drawn uniformly at random from all researchers in the network) for each unique team size present in the network. Then, for every team $p$ in the network, we take the difference of the observed value of $\tilde H_p$ and the average value $\mu^{(H)}_{\av{p}}$ from simulations of random teams of the same size, and divide by the standard deviation $\sigma^{(H)}_{\av{p}}$ of the results for the randomized teams. This gives us the $z$-score $z^{(H)}_p$ of the observed result $\tilde H_p$ in the null ensemble where researchers have no collaboration preferences, thus
\begin{align}
\label{zH}
z^{(H)}_p = \frac{\tilde H_{p}-\mu^{(H)}_{\av{p}}}{\sigma^{(H)}_{\av{p}}}.
\end{align}
For example, if a team of size $\av{p}=4$, we run 1,000 simulations drawing teams of $4$ at random from all authors in the network to get a vector $\vec{H}_{\av{p}}$ of simulation results, the mean and standard deviation of which we use in Eq. \ref{zH} as $\mu^{(H)}_{\av{p}}$ and $\sigma^{(H)}_{\av{p}}$ respectively. In the same manner, we compute a z-score $z^{(S)}_p$ using the same simulations, but take the measure of interest to be $\tilde S_{p}$ rather than $\tilde H_p$
\begin{align}
\label{zS}
z^{(S)}_p = \frac{\tilde S_{p}-\mu^{(S)}_{\av{p}}}{\sigma^{(S)}_{\av{p}}}.
\end{align}
We plot kernel density estimated probability densities of Eq. \ref{zH} and Eq. \ref{zS} for the full collaboration network in Figure \ref{group_fig}. We can see from these results that research teams tend to be composed of people with more homogeneous backgrounds than expected by chance, both with respect to primary discipline and full research profile. In particular, the distribution of $z^{(H)}_p$ has its mass centered at $z=-1$, indicating that most research teams have $\tilde H_p$ about one standard deviation lower (more concentrated) than expected on average for an uncorrelated random network. Additionally, the distribution of $z^{(S)}_p$ has its mass centered at $z=+1$, suggesting that many research teams have a $\tilde S_p$ about one standard deviation above (more similar discipline vectors $\vec{x}_i$) what is expected for a random team configuration. These results suggest that, in the interdisciplinary area, research teams as a whole tend to be composed of researchers with similar research backgrounds.

\subsection{Academic Performance and Collaboration Diversity}
\label{core_shell_section}
\begin{figure}
    \centering
    \includegraphics[width=.6\textwidth]{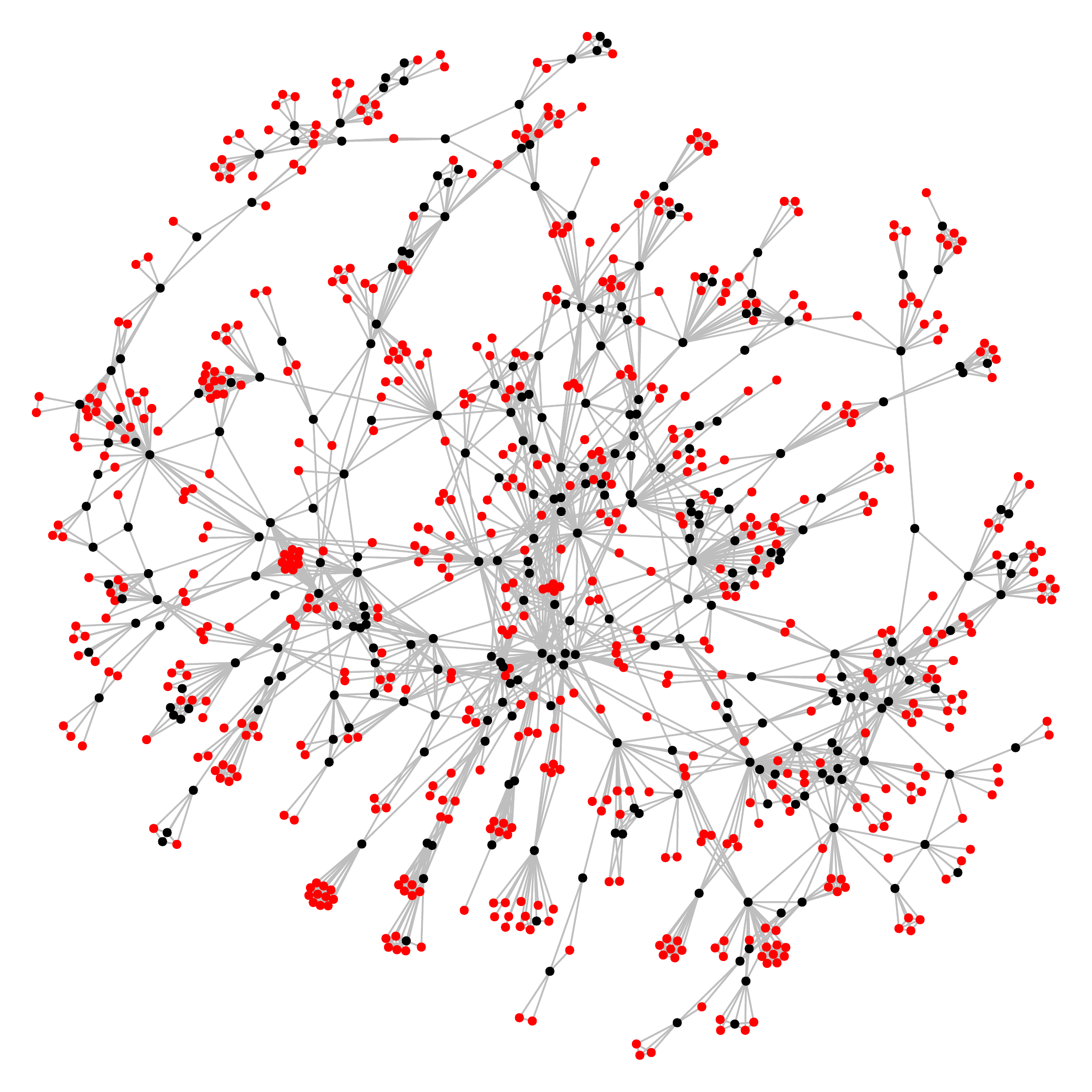}
    \caption{The giant component of the collaboration network used in this study, with nodes colored according to core-shell classification using the method from Section \ref{measures}. Core nodes are colored black, while shell nodes are colored red. This decomposition helps to separate the network into active and inactive collaborators.}
    \label{core_shell_pic}
\end{figure}

\begin{figure}
    \centering
    \includegraphics[width=.45\textwidth]{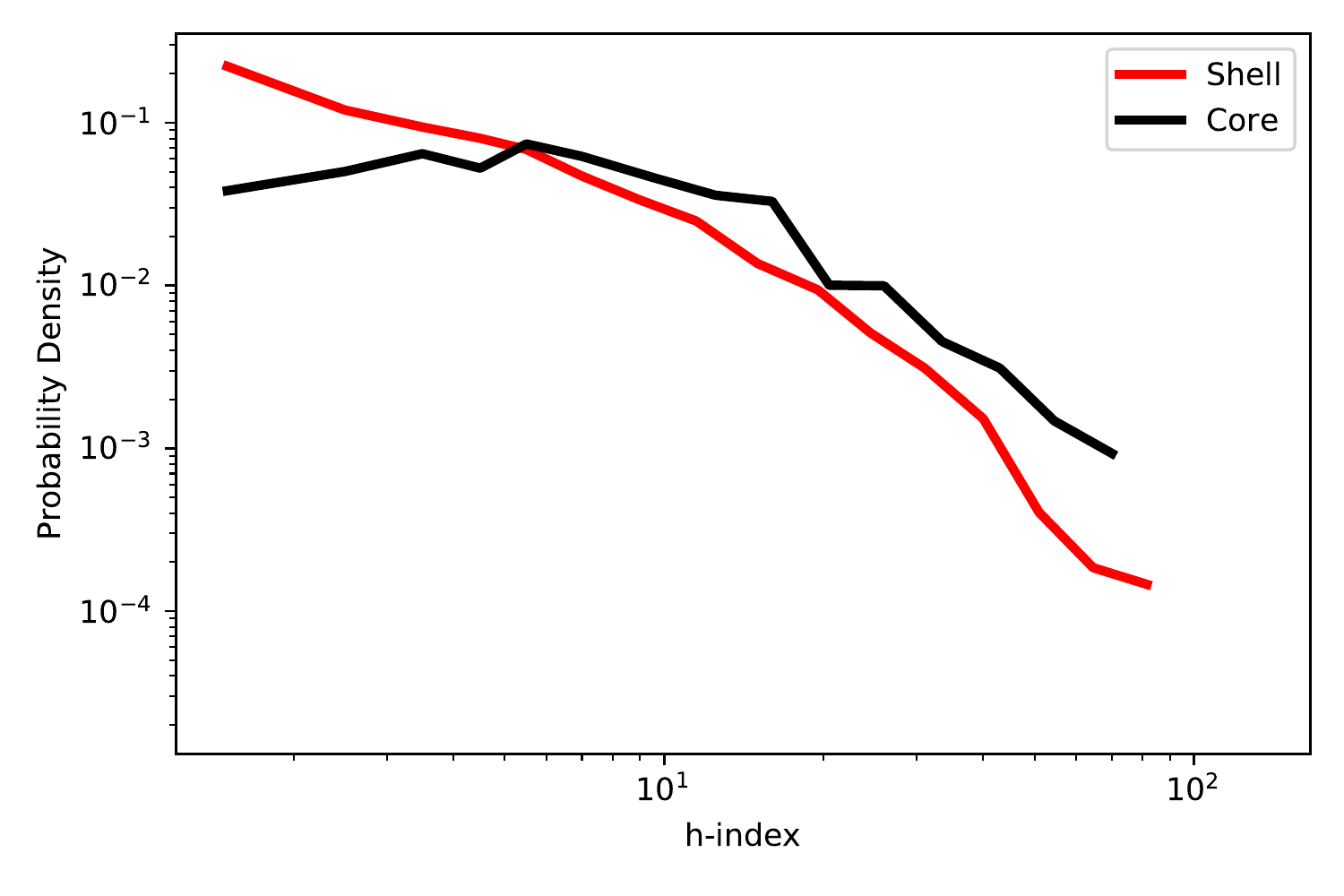}
    \includegraphics[width=.45\textwidth]{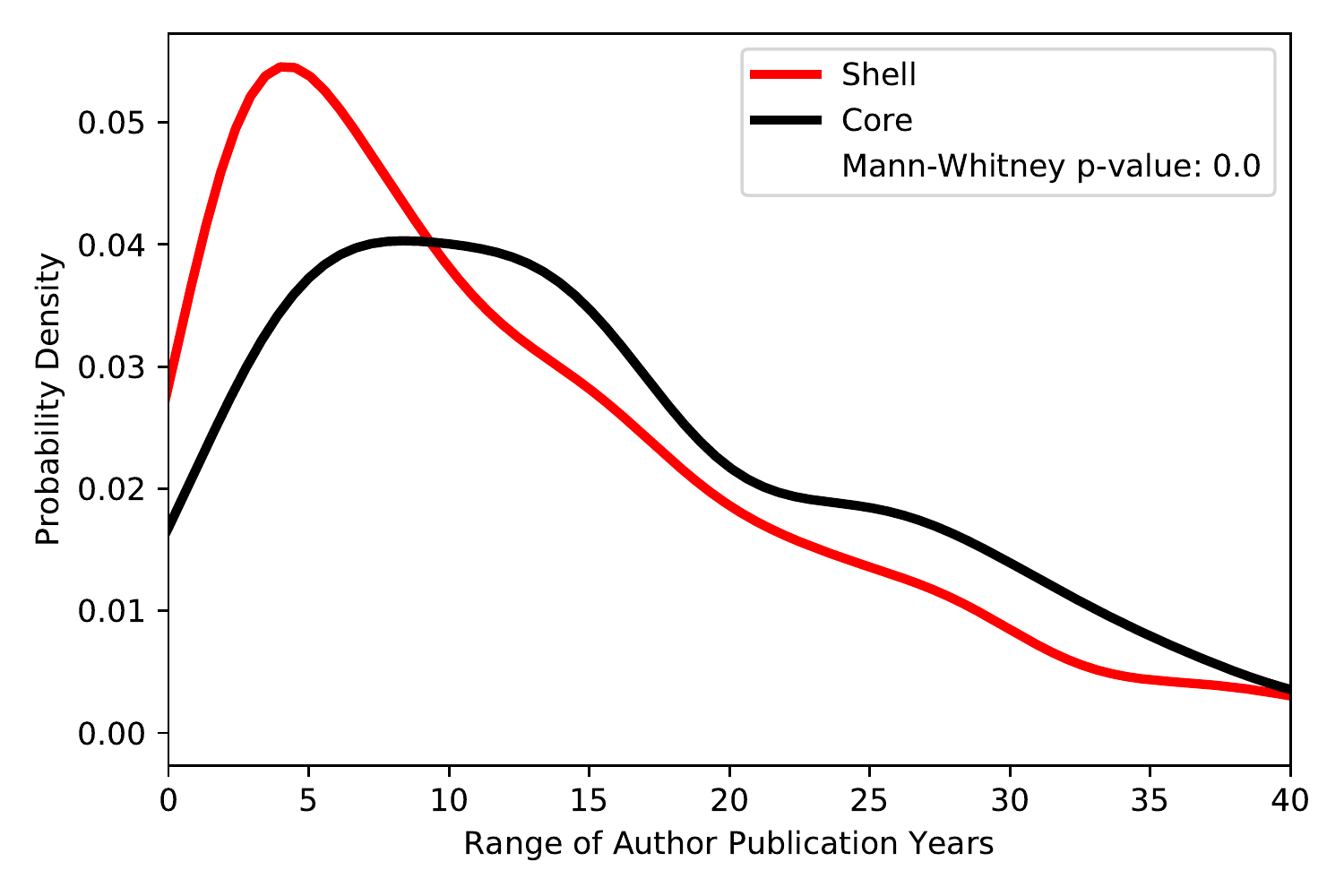}
    \caption{Distributions of h-index and years of publication experience for the core and shell in the decomposition shown in Fig. \ref{core_shell_pic}. The distributions for the core were significantly shifted above the distributions for the shell in both cases, indicating significantly more academic experience and success among the actively collaborating core researchers.}
    \label{h_and_years_fig}
\end{figure}

We apply the core-shell decomposition discussed in Section \ref{measures} to separate the active inter-group collaborators from the inactive ones, which is visualized in Figure \ref{core_shell_pic}. The decomposition reveals a shell of 1,602 nodes and a core of 420 nodes, indicating that most nodes in the network only participate in a single collaboration, and a smaller portion actively work with multiple groups. To examine the associations between structural diversity of authors in the collaboration network and academic experience and performance, we plot the distributions of h-index and years of publication experience (the difference between the earliest and latest publication on record for the author) for the core and shell nodes in Figure \ref{h_and_years_fig}. The results indicate that researchers in the core tend to have a systematically higher h-index and more publication experience than those in the shell. To statistically validate this claim, we apply both Mann-Whitney and Kolmogorov-Smirnov tests (as in Section \ref{core_shell_section}), finding that in all cases the results are statistically significant
(h-index: \text{Median 1}$=10$, \text{Median 2}$=4$, $n_1=420$, $n_2=1602$, $U \gg 10$ , $p \ll .01$, for one-tailed Mann-Whitney U test; $D=0.31$, $p\ll .01$ for KS test); (publication years: \text{Median 1}$=13$, \text{Median 2}$=9$, $n_1=420$, $n_2=1602$, $U \gg 10$ , $p \ll .01$, for one-tailed Mann-Whitney U test; $D=0.15$, $p\ll .01$ for KS test). These results suggest that, in the interdisciplinary area, the researchers who have longer working experience and better academic performance tend to be more active in collaborating with diverse groups on more projects. \\   

. 

\section{Discussion}
\subsection{Interdisciplinarity is mainly reflected by individual researchers in an interdisciplinary research area}

Our results suggest that interdisciplinarity is better demonstrated at the individual level than the dyadic or group level in an interdisciplinary research area. This implies that perhaps interdisciplinary research topics attract researchers who have experience in multiple fields, but this does not necessarily lead to diverse collaborations. Research experience in multiple fields strengthens the flexibility and adaptability of a researcher for engaging in projects that cross disciplinary boundaries. The capability of connecting knowledge across different disciplines also enables researchers to develop novel questions and analysis methods, which are central to interdisciplinary research. One potential challenge faced by interdisciplinary researchers is the competing demands of time and effort for each field they participate in. The findings of this study indicate that while interdisciplinary researchers involve in three to five disciplines, they can relatively equally contribute to all the fields they are interested in. However, the capacity to contribute to all fields equally is diminished as the number of research fields they participate in increases. Our findings on the prevalence of individually interdisciplinary researchers in this interdisciplinary research area highlight the importance of diverse training, which not only prepares individuals with a comprehensive knowledge base, but also supports them to collaborate in interdisciplinary fields.   \\

\subsection{Homophily is stronger than diversity for collaboration in interdisciplinary research}

Despite the presumed benefits for  collaboration with people from diverse academic backgrounds in interdisciplinary research areas, our study finds that researchers still prefer to collaborate with others who are alike in terms of their research background. Given that individual researchers tend to have interdisciplinary research background, we consider the multiple fields that individuals participate in while assessing pairwise similarity in the collaboration network, and we find that researchers prefer to collaborate with others who work in a similar set of fields. These findings indicate that homogeneity in pairwise collaborations is not constrained to the primary disciplines of individuals in interdisciplinary research, and that the interdisciplinary research experiences of individuals should be taken into consideration. Dyadic homogeneity and individual-level interdisciplinarity reduce transaction costs, ensure the diversity of the body of knowledge within a research group, and facilitate the development of a shared collaborative grounding. Our results may thus provide a theoretical contribution to understanding the development of collaboration in interdisciplinary research, as well as insight for reconsidering the definition of interdisciplinary research. A previous study  \cite{aboelela2007defining} considers the diversity of disciplines of researchers in a project as a dimension for defining interdisciplinary research. Based on the findings of this study, it is not necessary to have researchers from diverse disciplines in an interdisciplinary research, rather, interdisciplinarity can be reflected at the individual level instead of the group level. \\

\subsection{Diversity in collaborating with multiple groups is beneficial}

Based on the core-shell analysis of the collaboration network, we find that researchers who are active in collaborating with diverse groups on multiple projects tend to have a better academic performance and longer working years. This makes sense, as researchers with reputable track records and more experience have a greater pool of resources that facilitate the development of research collaborations with diverse groups on multiple projects. In a complementary way, collaborating with many teams on more projects can also enhance a researcher's academic performance, which unsurprisingly is positively associated with number of years publishing. However, confirming a causal relationship from this finding requires further research. 

\section{Conclusion}

This study proposes novel measures for assessing interdisciplinarity at three scales within research collaborations in an interdisciplinary area. Our findings contribute to the conceptual, theoretical, and methodological aspects of understanding research collaboration in interdisciplinary areas. \\

Firstly, we consider the discipline of researchers based on their publication records rather than departmental affiliations, which offers a means of defining and assessing the interdisciplinarity of individual researchers. Secondly, we introduce new measures for assessing interdisciplinarity at the individual, dyadic, and team levels which could be further employed in future studies on other datasets. Thirdly, a new cost-effective approach for identifying a core of nodes with diverse neighborhood structure in a network is proposed, which is especially effective on networks that are tree-like at the clique level, such as collaboration networks. In terms of theoretical contributions, this study strengthens our understanding of the underlying principles involved in developing collaborations in interdisciplinary research. Our results indicate that homophily is still an applicable principle for explaining collaborative relationships, and that additionally, individual interdisciplinarity and dyadic homogeneity together form the theoretical underpinnings of developing collaborations in interdisciplinary research. Lastly, the findings of the study shed light on the nature of team formation in practice, as well as highlight the importance of an interdisciplinary program. \\

It is important to support the development of interdisciplinary programs at both the institutional and national levels, as researchers with interdisciplinary backgrounds can better contribute to interdisciplinary collaboration. Regarding team formation in interdisciplinary research, it is important to consider the research experiences of individuals as well as the overlapping of individual interdisciplinarity among group members. Future studies are also needed to explore the factors affecting the success of research collaborations in interdisciplinary research areas.\\



\begin{thebibliography}{10}
\expandafter\ifx\csname url\endcsname\relax
  \def\url#1{\texttt{#1}}\fi
\expandafter\ifx\csname urlprefix\endcsname\relax\def\urlprefix{URL }\fi

\bibitem{van2015interdisciplinary}
Van Noorden, R. Interdisciplinary research by the numbers. Nature 525, 306-307 (2015).

\bibitem{clark1991grounding}
Clark, H. H. \& Brennan, S. E. Grounding in communication. (1991).

\bibitem{hertzum2008collaborative}
Hertzum, M. Collaborative information seeking: The combined activity of information seeking and collaborative grounding.
Information Processing \& Management 44, 957-962 (2008).

\bibitem{campbell2005overcoming}
Campbell, L. M. Overcoming obstacles to interdisciplinary research. Conservation Biology 19, 574-577 (2005).

\bibitem{lewin1948resolving}
Lewin, K. Resolving social conflicts; Selected papers on group dynamics. (1948).

\bibitem{ruef2003structure}
Ruef, M., Aldrich, H. E. \& Carter, N. M. The structure of founding teams: Homophily, strong ties, and isolation among
US entrepreneurs. American Sociological Review 195-222 (2003).

\bibitem{baggs1992association}
Baggs, J. G., Ryan, S. A., Phelps, C., Richeson, J. \& Johnson, J. The association between interdisciplinary collaboration
and patient outcomes in a medical intensive care unit. Heart \& Lung: The Journal of Critical Care 21, 18-24 (1992).

\bibitem{fewster2008interdisciplinary}
Fewster-Thuente, L. \& Velsor-Friedrich, B. Interdisciplinary collaboration for healthcare professionals. Nursing Adminis-
tration Quarterly 32, 40-48 (2008).

\bibitem{petri2010concept}
Petri, L. Concept analysis of interdisciplinary collaboration. In Nursing Forum, vol. 45, 73-82 (Wiley Online Library,
2010).

\bibitem{van2011factors}
Van Rijnsoever, F. J. \& Hessels, L. K. Factors associated with disciplinary and interdisciplinary research collaboration.
Research Policy 40, 463-472 (2011).

\bibitem{cummings2008collaborates}
Cummings, J. N. \& Kiesler, S. Who collaborates successfully? prior experience reduces collaboration barriers in distributed
interdisciplinary research. In Proceedings of the 2008 ACM Conference on Computer Supported Cooperative Work, 437-446
(2008).

\bibitem{guimera2005team}
Guimera, R., Uzzi, B., Spiro, J. \& Amaral, L. A. N. Team assembly mechanisms determine collaboration network structure
and team performance. Science 308, 697-702 (2005).

\bibitem{moody2004structure}
Moody, J. The structure of a social science collaboration network: Disciplinary cohesion from 1963 to 1999. American
Sociological Review 69, 213-238 (2004).

\bibitem{dahlander2013ties}
Dahlander, L. \& McFarland, D. A. Ties that last: Tie formation and persistence in research collaborations over time.
Administrative Science Quarterly 58, 69-110 (2013).

\bibitem{qin1997types}
Qin, J., Lancaster, F. W. \& Allen, B. Types and levels of collaboration in interdisciplinary research in the sciences. Journal
of the American Society for Information Science 48, 893-916 (1997).

\bibitem{kumar1985diversity}
Kumar Nayak, I. On diversity measures based on entropy functions. Communications in Statistics-Theory and Methods
14, 203-215 (1985).

\bibitem{aboelela2007defining}
Aboelela, S. W. et al. Defining interdisciplinary research: Conclusions from a critical review of the literature. Health
Services Research 42, 329-346 (2007).
  
\end{thebibliography}
\end{document}